\documentclass[%
 reprint,
superscriptaddress,
 amsmath,amssymb,
 aps,
pra,
]{revtex4-2}

\usepackage{float}
\usepackage{graphicx}% Include figure files
\usepackage{hyperref}
\usepackage{cleveref}
\usepackage{dcolumn}% Align table columns on decimal point
\usepackage{bm}% bold math
\usepackage{xcolor}
\usepackage{yhmath,scalerel}
\usepackage{gensymb}
\usepackage{amsmath}
\usepackage{amssymb}

\newcommand{\ket}[1]{\ensuremath{|{#1}\rangle}}

\newcommand{\mic}{\,$\mu$m\:}

\DeclareMathAlphabet\mathbfcal{OMS}{cmsy}{b}{n}

\begin{document}

\preprint{APS/123-QED}

\title{Sub-Doppler cooling of a trapped ion in a phase-stable polarization gradient}

\author{Ethan Clements}
\email{eclement@mit.edu}
\author{Felix W. Knollmann}
\author{Sabrina Corsetti}
\author{Zhaoyi Li}
\author{Ashton Hattori}
\author{Milica Notaros}
\affiliation{%
Massachusetts Institute of Technology, Cambridge, Massachusetts 02139, USA
}
\author{Reuel Swint}
\affiliation{
Lincoln Laboratory, Massachusetts Institute of Technology, Lexington, Massachusetts 02420, USA
}
\author{Tal Sneh}
\affiliation{%
Massachusetts Institute of Technology, Cambridge, Massachusetts 02139, USA
}
\author{May E. Kim}
\affiliation{
Lincoln Laboratory, Massachusetts Institute of Technology, Lexington, Massachusetts 02420, USA
}
\author{Aaron D. Leu}
\affiliation{Clarendon Laboratory, Department of Physics, University of Oxford, Parks Road, Oxford OX1 3PU, United Kingdom}
\author{Patrick Callahan}
\author{Thomas Mahony}
\affiliation{
Lincoln Laboratory, Massachusetts Institute of Technology, Lexington, Massachusetts 02420, USA
}
\author{Gavin N. West}
\affiliation{%
Massachusetts Institute of Technology, Cambridge, Massachusetts 02139, USA
}
\author{Cheryl Sorace-Agaskar}
\author{Dave Kharas}
\author{Robert McConnell}
\author{Colin D. Bruzewicz}
\affiliation{
Lincoln Laboratory, Massachusetts Institute of Technology, Lexington, Massachusetts 02420, USA
}
\author{Isaac L. Chuang}
\author{Jelena Notaros}
\affiliation{%
Massachusetts Institute of Technology, Cambridge, Massachusetts 02139, USA
}
\author{John Chiaverini}\affiliation{%
Massachusetts Institute of Technology, Cambridge, Massachusetts 02139, USA
}%
\affiliation{
Lincoln Laboratory, Massachusetts Institute of Technology, Lexington, Massachusetts 02420, USA
}

\date{\today}% It is always \today, today,
             %  but any date may be explicitly specified

\begin{abstract}
Trapped ions provide a highly controlled platform for quantum sensors, clocks, simulators, and computers, all of which depend on cooling ions close to their motional ground state. Existing methods like Doppler, resolved sideband, and dark resonance cooling balance trade-offs between the final temperature and cooling rate. A traveling polarization gradient has been shown to cool multiple modes quickly and in parallel, but utilizing a stable polarization gradient can achieve lower ion energies, while also allowing more tailorable light-matter interactions in general. In this paper, we demonstrate cooling of a trapped ion below the Doppler limit using a phase-stable polarization gradient created using trap-integrated photonic devices. At an axial frequency of $2\pi\cdot1.45~ \rm MHz$ we achieve $\langle n \rangle = 1.3 \pm 1.1$ in $500~\mu \rm s$ and  cooling rates of ${\sim}0.3 \, \rm quanta/\mu s$. We examine ion dynamics under different polarization gradient phases, detunings, and intensities, showing reasonable agreement between experimental results and a simple model. Cooling is fast and power-efficient, with improved performance compared to simulated operation under the corresponding running wave configuration.

\end{abstract}

\maketitle
 
A fundamental step for quantum sensors and computers using trapped ions is cooling into the motional ground state because it improves the fidelities of quantum gates~\cite{bermudez2017assessing}. Current systems built for trapped-ion based quantum information processing rely mainly on resolved-sideband cooling to prepare ion crystals close to the motional ground state~\cite{qscout, pino2021demonstration, oxfordionics}. Sideband cooling requires resolving and addressing individual motion-subtracting sidebands on narrow-linewidth electronic transitions to remove a quantum of motion from a particular mode. Since the ion must be driven on a forbidden transition or via a Raman transition, this process can be power intensive. Further, since each mode of motion must be individually addressed, this process is slow and comprises a large percentage of the computational cycle time~\cite{pino2021demonstration}. Sideband cooling will still be necessary to reliably prepare trapped-ion systems in their motional ground state but complementary sub-Doppler cooling methods can be incorporated to reduce the average power consumption and time overhead of cooling routines.

In general, the cooling and trapping of atoms has enabled an entire field of physics with results spanning from Bose-Einstein condensates~\cite{bec} to exquisite control of individual quantum systems~\cite{wineland1998experimental}. A central technique in this development has been Sisyphus cooling~\cite{cirac1992laser,wineland1992sisyphus,cirac1993laser,petra2003numerical, phatak2024generalized,kazantsev1985kinetic,dalibard1985dressed}, which originated from the surprising observation that thermal profiles of atoms in the first magneto-optical traps displayed temperatures below the Doppler cooling limit~\cite{lett1988observation,lett1989atoms}. A version of Sisyphus cooling called polarization gradient cooling (PGC) offers a potentially broadband and parallel way to cool many motional modes close to the ground state faster than sideband cooling. This cooling mechanism was theoretically explored in both a phase stable polarization gradient and a running wave polarization gradient~\cite{cirac1993laser}. In the running-wave approach, demonstrated in recent experiments~\cite{ejtemaee20173d,joshi2020polarization,li2022robust}, the relative frequency of intersecting light fields is varied to create a spatially moving gradient of polarization. The spatial dependence of the gradient is averaged over, increasing the steady-state temperature by a factor of two~\cite{joshi2020polarization}.

In this letter, we demonstrate sub-Doppler cooling in a phase stable polarization gradient enabled by the interferometric stability obtained from on-chip splitting of light using trap-integrated photonic light delivery~\cite{vasquez2023control,corsetti_pgc}. To our knowledge, this work represents the first experimental cooling of a single atom or ion in a phase-stable polarization gradient. Using integrated optical beams, we demonstrate phase-dependent dynamics of the ion motion and characterize the cooling as a function of relevant parameters; we compare the results to values obtained via both a simple analytical and a more accurate numerical model.  We obtain an average axial mode occupation of $\langle n \rangle = 1.3 \pm 1.1$ in $500~\mu$s for a motional frequency of $2\pi\cdot1.45~\rm MHz$ and obtain sub-Doppler cooling to a few quanta within ${\sim} 50$ $\mu$s. We note that additional details about device design, device characterization, and further ion-to-polarization-gradient stability characterization are detailed in an adjoined publication~\cite{corsetti_pgc}.
\nocite{mehta2016integrated,agaskar2019versatile, niffenegger2020integrated,corsetti_pgc}
%%%%% End of intro %%%%%%%

\begin{figure*}[ht]
\centering
\includegraphics[width=.90\textwidth]{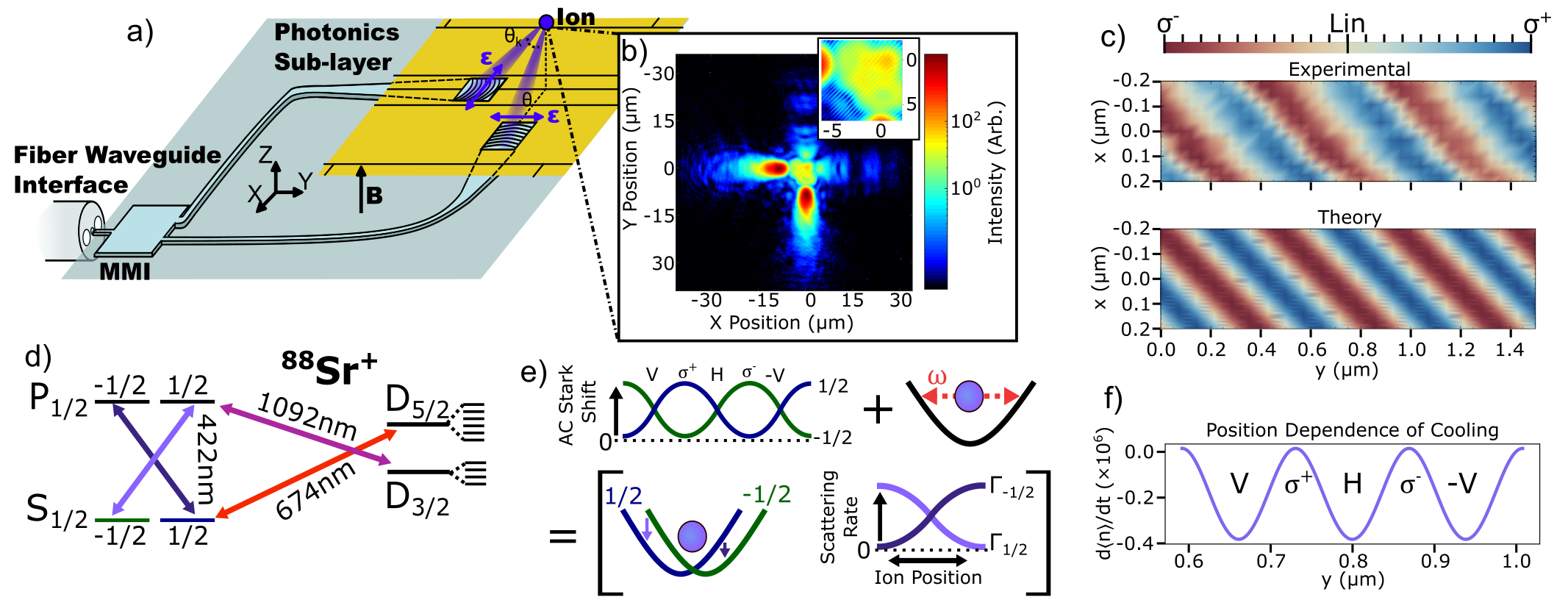}
\caption{\label{fig:overview}  Experimental setup, polarization mapping, and schematic of cooling mechanism. (a) Schematic of the trap layout used for polarization gradient cooling: the trap electrodes are gold and the simplified photonic structures are blue. Also shown are the grating emission angle $\theta$, angle between the k-vectors of the two beams $\theta_k$, emitted polarizations $\epsilon$, coordinate system, and quantization field vector B. (b) A cross-section of the grating emission imaged with a high-NA objective shows the  overlap of the two beams at the ion location (plot origin). Inset of the cross-section plot shows the intensity structure of the laser light at the ion location in detail. We observe a periodic variation in intensity from interference of the H-fields of the two beams. The intensity variation is a secondary effect to the polarization gradient in the cooling process. We further detail the cause of this intensity structure in the supplement~\cite{supplement}. At the intersection of the two emitted lasers we measure the spatial polarization dependence (c) by using the polarization gradient light for state preparation and using electron shelving to map the population to a fluorescence measurement.  At the nodes, the polarization is linear with equal components of $\sigma^{+}$ and $\sigma^{-}$. The theoretical profile at the bottom of (c) is the state preparation outcome for a point particle assuming interference of plane-waves with direction and intensity comparable to experimental values. In (d) we outline the electronic level structure for ${}^{88}\rm{Sr}^+$ with transition wavelengths labeled for laser light used in this experiment, the two Zeeman states in both S\(_{1/2}\) and P\(_{1/2}\) used for cooling, and Zeeman sub-states of the two metastable excited states. (e) Simplified schematic of how cooling occurs in a polarization gradient. The $S_{1/2}$ Zeeman ground states are AC stark shifted spatially. The ion is localized at a node of the polarization gradient in the harmonic trapping potential with characteristic trapping frequency $\omega$. These shifts, combined with the spatial dependence of the scattering rate, result in a reduction of motional energy of the ion following photon scattering. (f) We use an analytical model to estimate the spatial dependence of the cooling rate at $\textrm{x} = 0$ and label the polarization at different spatial positions.}
\end{figure*}

In our demonstration of PGC, we trap a singly-ionized strontium ion ${\sim}$50\,\mic over the metal layer of a surface-electrode RF Paul trap. The trap is defined by niobium electrodes that are patterned over alumina and silicon nitride waveguides clad in silicon dioxide; details can be found in refs.~\cite{mehta2016integrated,agaskar2019versatile, niffenegger2020integrated,corsetti_pgc}. Light from a 422~nm laser stabilized using a Rb vapor-cell~\cite{mausezahl2024tutorial} is sent to the ion-trap chip using a polarization-maintaining fiber routed from outside of the vacuum system to an alumina waveguide facet at the edge of the trap chip. On chip, the light is equally split by an alumina multi-mode interferometer (MMI)~\cite{soldano1995optical}. Each arm is routed to a grating emitter through a combination of alumina and silicon-nitride waveguides bridged by vertical layer transitions ~\cite{agaskar2019versatile}. The dual-layer silicon-nitride gratings \cite{corsetti2023integrated,corsetti_pgc} used in these experiments are designed to emit focused beams with a wavevector angled $45^{\circ}$ from the trap surface and light linearly-polarized parallel to the surface of the chip. In practice, the fabricated gratings emit at an angle of $49.42 \pm 0.58^{\circ}$ with respect to the surface of the chip, and the $1/e^2$ beam waist at the beam intersection is $\sim$3~$\mu$m. The gratings are positioned perpendicular to each other with one grating emitting along the linear trap axis. These device details are highlighted in Fig.~\ref{fig:overview}(a) and (b) and further detailed in ref.~\cite{corsetti_pgc}. The emitted laser light creates a modified lin$\perp$lin polarization gradient as shown in Fig.~\ref{fig:overview}(c). In this experiment, fabrication biases resulted in the primary intensity portion of the beam from each grating overlapping $\sim$10$~\mu$m above the designed ion height, outside of the stable trapping volume. The results presented here use a polarization gradient in an overlap of lower intensity portions of the beams at the design height of the ion. The imperfections arising from this bias are a minor limitation of this work but can be straightforwardly addressed in future implementations. All other light used in this work is delivered via free-space beams focused onto the ion by optics external to the vacuum system but could be integrated as well~\cite{mehta2016integrated,niffenegger2020integrated}.

We map the polarization gradient in our system by measuring the population in the $\text{S}_{1/2}$ Zeeman states as we vary the ion position along the x and y directions. At different ion displacements, the polarization varies between circular and linear polarization, affecting which states are addressed. We apply a 10~$\mu$s pulse of the polarization gradient light with a saturation parameter $s \ll 1$ and then use electron-shelving~\cite{e_shelving} to the metastable $\text{D}_{5/2}$ state to measure the imbalance between the $m_{j} = \pm 1/2$ ground states using fluorescence detection. The saturation parameter is defined as \(s=\frac{\Omega^2/2}{\delta^2 + \Gamma^2/4}\), where \(\Omega\) is the Rabi frequency of the dipole transition, $\delta$ is the angular laser detuning, and $\Gamma$ is the transition linewidth $(2\pi\cdot20.2 \rm\,MHz$). We calibrate the laser intensity by measuring the light shift of the $\ket{S_{1/2}, m_j = \pm 1/2}\leftrightarrow \ket{D_{5/2}, m_j = \pm 5/2}$ electric quadrupole transitions~\cite{supplement} (see Fig.~\ref{fig:overview}(d) for Sr$^{+}$ level diagram). This method is similar to other demonstrations of polarization mapping~\cite{schaetz2024}, however in this case we can make a measurement of the gradient without stroboscopically sampling the data because of the inherent phase stability of the light field. An example spatial scan of the polarization and an ideal theoretical comparison are shown in Fig.~\ref{fig:overview}(c). In these measurements we observe an upper bound of 70$\%$ contrast for the population fringes, limited primarily by the thermal state of the ion motion (the polarization purity and relative beam intensities additionally contribute to a smaller degree). In a simplified picture the temperature dependence of the contrast comes from the thermal state of the ion sampling a finite section of the polarization gradient. This causes the polarization to be spatially averaged and limits state preparation due to other polarization components pumping population out of the target state (details in supplementary material~\cite{supplement}).

\begin{figure}[!htbp]
\centering
\includegraphics[width=.45\textwidth]{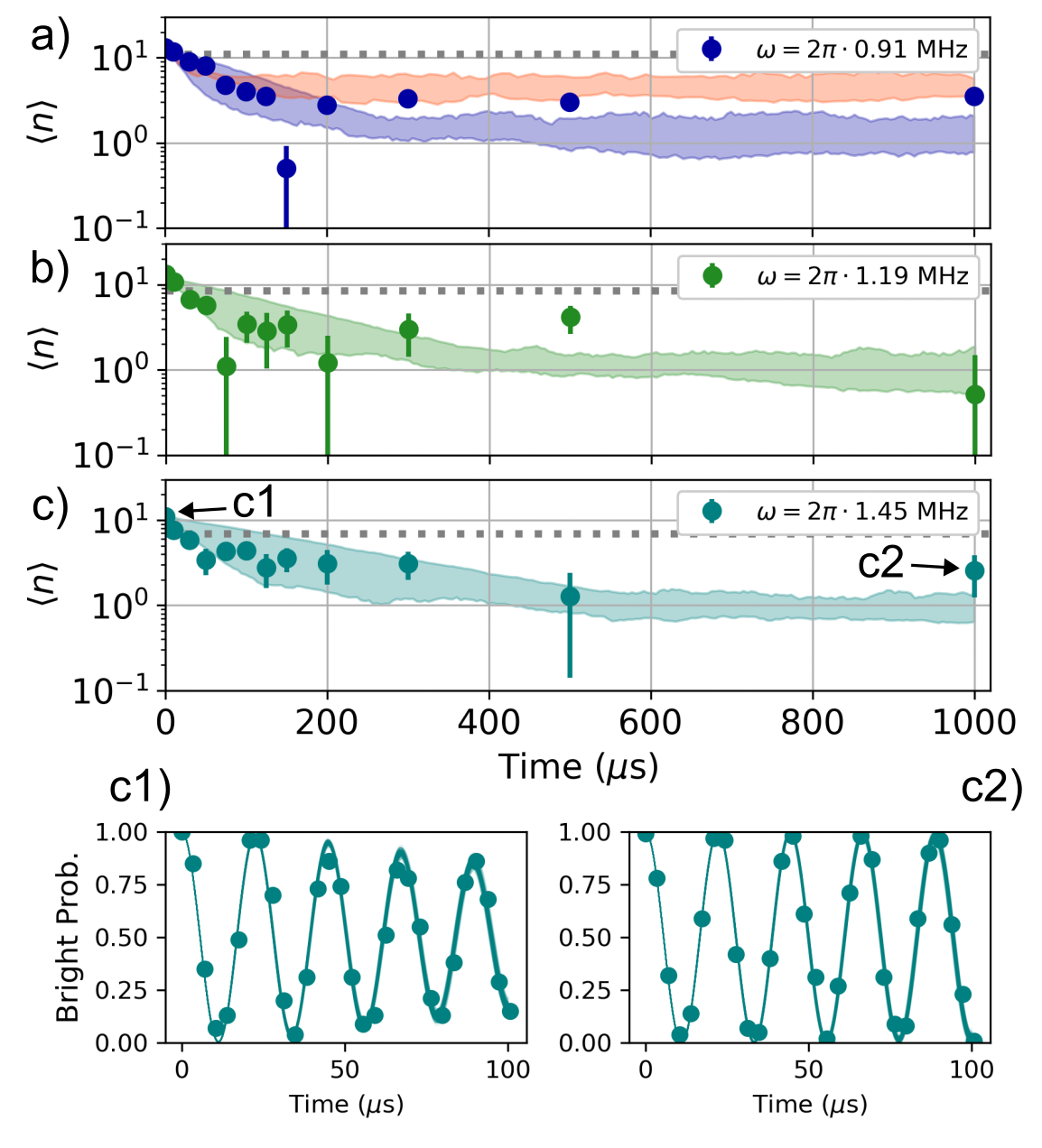}
\caption{\label{fig:occupation_v_time} Experimentally measured and simulated time dynamics of polarization gradient cooling. The average motional excitation is plotted as a function of applied PGC duration to show sub-Doppler cooling for different axial trap frequencies, (a), (b), and (c). As the axial frequency is varied the resonant saturation parameter is held constant to $s_0 \approx 4$ and the detuning from resonance is set to $2\pi\cdot140$ MHz. The gray dashed line is the theoretical Doppler cooling limit ($\Gamma/2\omega$). The simulated time dynamics assuming saturation parameters in the range of 2.5-6.5 are plotted as filled regions to account for possible uncertainties in the laser intensity (from simulations we expect that cooling with a higher resonant saturation parameter yields a shorter cooling time, but higher steady state motional occupation). The filled red data is simulated data of running-wave PGC using identical experimental parameters and a relative laser detuning of $2\pi\cdot100$~kHz. This shows the expected factor of two difference in steady-state Fock state occupation. Sub-figures (c1) and (c2) show the carrier Rabi-time scan data and fits used to determine the temperature of the ion before PGC is applied and after 1~ms of PGC.  }
\end{figure}
\begin{figure*}[htbp]
\centering
\includegraphics[width=.80\textwidth]{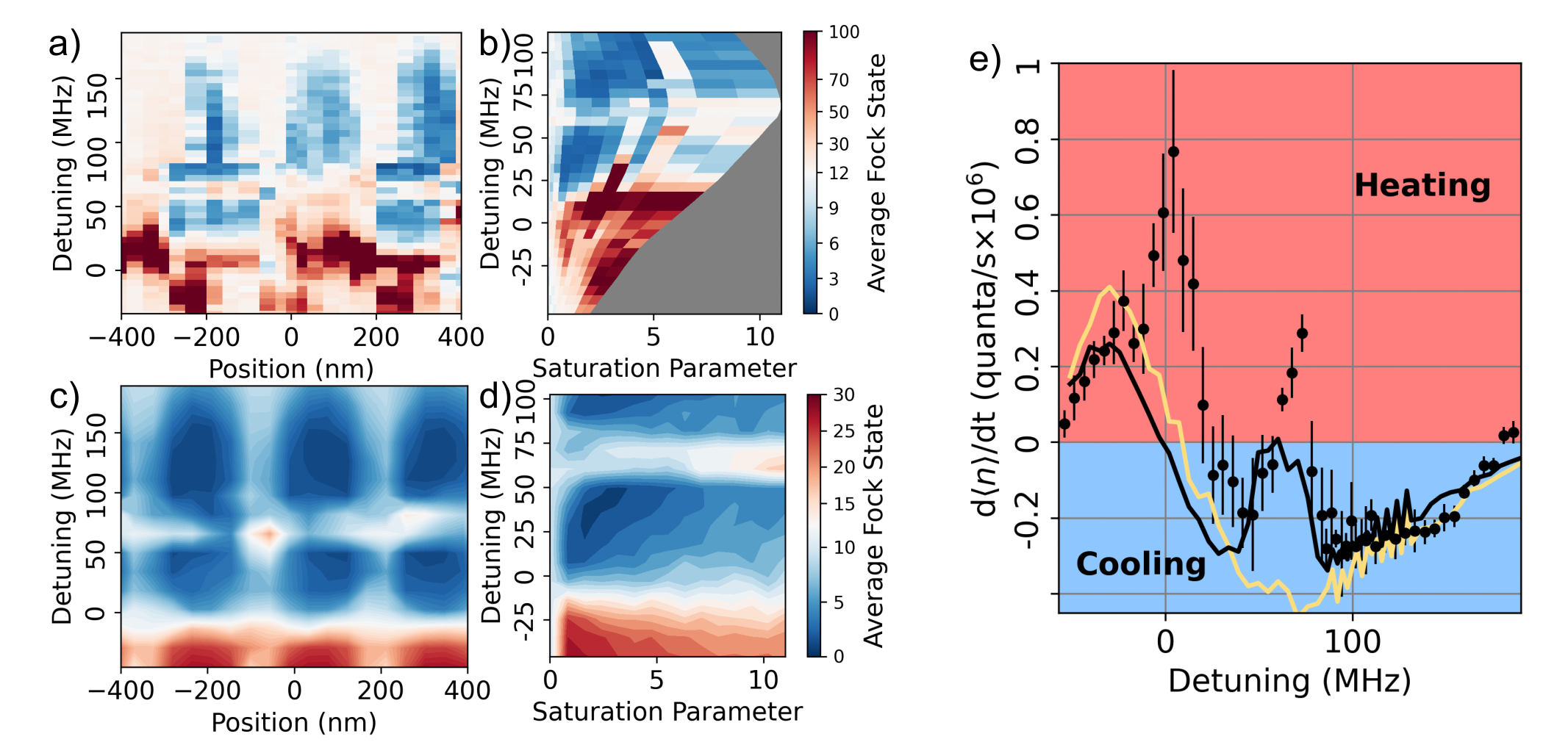}
\caption{\label{fig:param_space} Parameter space mapping of polarization gradient cooling (experiment and simulation). We experimentally determine the position, intensity, and frequency dependence of PGC and compare it to numerical simulations. Data and simulations in this figure use an axial frequency of $\omega = 2\pi\cdot0.91~\rm MHz$. The resonant saturation parameter between \(\delta\) = $2\pi\cdot50$ MHz and $2\pi\cdot150$ MHz is $s_0 \sim 4$ but outside of this range the intensity was limited by the diffraction efficiency of the acousto-optic modulator (AOM) used for frequency tuning (see supplement for details~\cite{supplement}). (a) As the laser detuning and the position of the ion along the axial trapping direction are varied, the average motional Fock state occupation following 200~$\mu$s of PGC displays characteristic heating and cooling regions. (b) Intensity dependence of the cooling rate. The envelope of the data is a result of the frequency dependence of the AOM diffraction efficiency. Subfigures (c) and (d) show numerical results for the average motional Fock state occupation, after 200~$\mu \rm s$ of cooling is applied, as the laser detuning and position (c) and the detuning and saturation parameter (d) are varied. Simulations used the Monte-Carlo solver in QuTip and accounted for the saturation parameter variation as the frequency was varied and had an upper Fock state limit of 30. (e) Cooling rate as a function of detuning at the ion position where the $\sigma^+/\sigma^-$ polarization components are balanced. These rates are determined from linear fits to the motional occupation at short cooling times. (Yellow) Simulated cooling rate considering only the Zeeman manifolds of the $S_{1/2}$ and $P_{1/2}$ electronic states. This model does not reproduce the feature seen at $\sim 2\pi\cdot75$~MHz. (Black) Simulated cooling rate with the addition of the Zeeman manifold of the $D_{3/2}$ state which produces the feature at $\sim 2\pi\cdot75$~MHz, indicating that this feature is a result of coherences between the $S_{1/2}$ the D\(_{3/2}\) state.
}
\end{figure*}

We use ion-based polarization mapping to characterize the gradient stability and spatially position the ion in the polarization gradient for cooling experiments. When running the experiments, the relative position between the ion and polarization gradient drifted by $\sim$10 nm/hour due to variation of the static electric field background. This drift was minimized using ion measurements to feed back on and stabilize the relative position as detailed in the supplemental material~\cite{supplement}.

The theory for polarization gradient cooling of a trapped ion has been studied for \(J=1/2\) to \(J'=3/2\)~\cite{cirac1993laser} and \(J=1/2\) to \(J'=1/2\) transitions~\cite{joshi2020polarization}. This prior work assumes counter-propagating beams of orthogonal linear polarizations (lin\(\perp\)lin) with the magnetic field aligned along the axis of propagation. This orientation creates a polarization gradient that transitions between circular and linear polarization with a theoretical period of $\lambda/2$ along the primary axis. The cooling mechanism can be summarized as follows and is depicted in Fig.~\ref{fig:overview}(e): due to angular momentum selection rules, the $\sigma$ polarized light creates state-dependent light shifts of the ground state that modify the bare trapping potential. Spatially dependent optical pumping by the $\sigma^{\pm}$ polarized light into the electronic state with the lower modified harmonic potential reduces the kinetic energy of the ions. For a $J=1/2$ to $J'=1/2$ transition, synchronizing the direction of the light-shifted potential with the position-dependent scattering rate requires the polarization gradient light to be blue-detuned.

Our setup differs from the canonical lin$\perp$lin case; we maintain mutual orthogonality of the magnetic field to the linear polarizations of the two beams, but without counter-propagation. This creates a phase-stable polarization gradient of $\sigma^{+/-}$ in the $\hat{x}+\hat{y}$ direction, which overlaps with all three modes of the ion’s motion (see Fig.~\ref{fig:overview}(c)). The period of the polarization gradient is scaled by a geometric factor of $1/\cos(\theta_k)$ from the canonical case (this system is designed to have an angle between the two beams of $\theta_{k}=$ 60$^{\circ}$). It is therefore convenient to use an effective wavelength $\lambda_{\rm eff} = \lambda \cos(\theta_{\rm ax})/\cos(\theta_k)$, where $\theta_{\rm ax}$ is the angle between the axis of interest and the primary polarization gradient axis. For $\theta_k = 60^{\circ}$ and $\theta_{\rm ax} = 45^{\circ}$ we expect $\lambda_{\rm eff} = 597 \rm ~nm$ but due to beam misalignment we observe a $\lambda_{\rm eff}$ of $555 \rm ~nm$ (from the polarization map measurements, we estimate errors in these angles of 2.2$^{\circ}$ and 1.1$^{\circ}$ for $\theta_{\rm ax}$ and $\theta_{k}$, respectively). The experimentally measured $\lambda_{\rm eff}$ is used in modeling and analysis.

Using a semiclassical model~\cite{cirac1993laser} one can calculate the cooling rate $W(\phi)$ as the ion position is varied in the polarization gradient to be:
\begin{equation}
W(\phi) = \frac{16}{9} \frac{\eta^2 \Gamma \delta s_0^2}{3\omega(1+4\delta^2/\Gamma^2)} \cos^2 (2\phi).
\end{equation} Where $\eta$ is the Lamb-Dicke parameter, $\Gamma$ is the cooling transition linewidth, $\delta$ is the laser detuning, $\omega$ is the motional frequency, $\phi$ is the relative phase of the two lasers used to describe the ion position, and $s_0$ is the on resonance saturation parameter $s(\delta = 0)$. This equation and other details of the semi-classical model for this system are further described in the supplement~\cite{supplement}. A plot of the spatial dependence of cooling is shown in Fig.~\ref{fig:overview}(f) where $s_0 = 4$, $\delta = 2\pi\cdot140 \rm ~MHz$, $\Gamma = 2\pi\cdot 20.2 \rm ~MHz$, $\omega = 2\pi\cdot1 \rm ~MHz$, and $\eta$ is set by $\omega$ and the laser wavelength of 422~nm. This model provides intuition for the rate of cooling and spatial dependence but is limited in its ability to capture observed experimental characteristics. 

As a result, we also numerically model this system. We compare our experimental results to simulations using a Lindblad master equation solver in QuTip~\cite{johansson2012qutip}. We build our model using the simplifying assumption of two counter-propagating waves along the $z$ direction with orthogonal polarizations and scale the wavenumber $k$ corresponding to the projection of the polarization gradient onto the axial trap axis. The Hamiltonian used in simulation includes terms for the harmonic trapping potential and spin-motion coupling caused by dipole transitions driven by the polarization gradient. We also include coupling between the $P_{1/2}$ and $D_{3/2}$ manifolds to characterize the impact of dark resonances~\cite{allcock2016dark} on the PGC dynamics. For simplicity we only model dynamics along the axial trapping direction. Details of the model and resulting simulated data are included in the supplementary material~\cite{supplement}.

Figure~\ref{fig:occupation_v_time} shows sub-Doppler cooling of the axial mode using PGC. We experimentally probe the dynamics by fitting to carrier Rabi oscillations on the $\ket{\text{S}_{1/2},-\frac{1}{2}}\rightarrow\ket{\text{D}_{5/2},-\frac{5}{2}}$ transition and extracting the steady-state motional occupation $\langle n_{ss} \rangle$ assuming a thermal distribution of the axial Fock states (the probe laser is aligned along the axial direction). We plot the average motional state occupation as a function of the duration of the applied polarization-gradient light pulse after an initial 1 ms of Doppler cooling. The experimental time series is in reasonable quantitative agreement with simulated data for both cooling rate and steady state temperature. We compare the cooling results for $\omega = 2\pi\cdot0.91$ MHz against simulations of cooling at this motional frequency with a running wave and observe a lower steady state temperature in the phase stable case. All modeling is further discussed in the supplement~\cite{supplement}.

We leverage the phase stability of our polarization gradient to explore the cooling parameter space by measuring $\langle n_{ss} \rangle$ as we vary laser power, laser frequency, and the position of the ion in the polarization gradient (Fig.~\ref{fig:param_space}(a) and (b)). We observe the transition from heating to cooling as we shift the frequency of the cooling laser further blue of resonance. All of these measurements were taken with an axial motional frequency of $2\pi\cdot0.91~\text{MHz}$ and a 200~$\mu$s pulse of PGC after initial Doppler cooling. The bands of cooling have a spatial period that is in agreement with the measured and theoretical spatial periods of the polarization gradient and coincide with the locations where the $\sigma^+/\sigma^-$ polarization components are balanced. The asymmetry seen between the cooling bands is likely caused by spatial variation of the laser intensity due to the imperfect overlap of the two beams and the interference of non-orthogonal polarization and H-field components of the light (see inset of Fig.~\ref{fig:overview}(b) and supplement for details~\cite{supplement}).

We also determine the heating and cooling rate as the detuning from resonance of the 422~nm laser is varied when the ion is placed at balanced $\sigma^+/\sigma^-$ polarization [Fig.~\ref{fig:param_space}(e)] by using a linear fit to $\langle n \rangle$ at short cooling times. A numerical simulation of the dynamics [presented in Fig.~\ref{fig:param_space}(c) and (d)] quantitatively agrees with the experimental results and captures the heating peak at ${\sim} 2\pi\cdot70$~MHz caused by coherences between the S\(_{1/2}\) and D\(_{3/2}\) state, which is visible across all plots in Fig.~\ref{fig:param_space}, when using a detuning from resonance of $2\pi\cdot140$~MHz and a repumper saturation parameter of 5000. 

We have demonstrated and characterized sub-Doppler cooling using a phase-stable polarization gradient enabled by trap-integrated photonics. Our experimental results both quantitatively and qualitatively agree with our numerical predictions. We observe fast cooling to substantially below the Doppler limit: $< 50$ $\mu$s to a few motional quanta. As we use \(s_0 < 5\) on a dipole-allowed transition, this cooling method is very laser power efficient. For similar conditions, sideband cooling on the quadrupole transition to the same temperature would require 4 orders of magnitude more laser power (for an equivalent \(\sim\)5~$\mu$m spot size) and 1 ms of cooling time. Moreover, we observe faster cooling rates and lower steady-state temperature for phase-stable PGC when compared to simulation of cooling with a running wave gradient (See supplement~\cite{supplement}). Further improvements could be made to the steady state temperature and cooling rate by changing the repumper frequency or making it incoherent to destabilize the dark resonances~\cite{lindvall2013unpolarized} and by increasing the laser intensity and detuning. Our models show that one could also adaptively change the saturation parameter or detuning during the cooling pulse to minimize the time needed to reach an optimal steady state temperature. 

In future work, one could characterize the cooling rate and steady state temperature across multiple radial and axial modes in longer ion chains. The interplay between the ion motional coupling and the spatial structure of the polarization gradient could lead to interesting Fock state dynamics as the spatial extent of the ions reaches the length scale of the gradient period. Relatedly, the precise spatial control over the light matter interaction afforded by the phase stable intensity or polarization gradient could allow for the study of frictional forces at the few atom scale and cooling dynamics of complex, multilevel systems~\cite{bonetti2021quantum,timm2021quantum,cheuk2018lambda,mccarron2018laser}.
\begin{acknowledgments}
We thank Alfredo Ricci Vásquez for comments on the document. This work is supported by a collaboration between the US DOE and other Agencies. This material is based upon work supported by the U.S. Department of Energy, Office of Science, National Quantum Information Science Research Centers, Quantum Systems Accelerator. Additional support is acknowledged from the NSF Quantum Leap Challenge Institute Hybrid Quantum Architectures and Networks (QLCI HQAN) (2016136), NSF Quantum Leap Challenge Institute Quantum Systems through Entangled Science and Engineering (QLCI Q-SEnSE) (2016244), MIT Center for Quantum Engineering (H98230-19-C-0292), NSF Graduate Research Fellowships Program (GRFP) (1122374), Department of Defense National Defense Science and Engineering Graduate (NDSEG) Fellowship, MIT Rolf G. Locher Endowed Fellowship, and MIT Frederick and Barbara Cronin Fellowship. This material is based upon work supported by the Department of Energy and the Under Secretary of Defense for Research and Engineering under Air Force Contract no. FA8702-15-D-0001. Any opinions, findings, conclusions, or recommendations expressed in this material are those of the author(s) and do not necessarily reflect the views of the Department of Energy or the Under Secretary of Defense for Research and Engineering. ILC acknowledges support by the NSF Center for Ultracold Atoms. EC was supported by an appointment to the Intelligence Community Postdoctoral Research Fellowship Program at MIT administered by Oak Ridge Institute for Science and Education through an interagency agreement between the U.S. Department of Energy and the Office of the Director of National Intelligence.

\paragraph*{Author Contributions}
The photonic structures were designed by SC, AH, MN, RS, and TS. The trap was laid out and fabricated by PC, TM, and DK and assembled by MEK, EC, and FWK. The experiments with ions were conducted by EC, FWK, SC, and ADL. The photonics tests were done by SC, AH, FWK, GNW, and MN. The theoretical model was made by EC and ZL. RM, CB, CSA, ILC, JN, and JC advised the work and acquired funding.
\end{acknowledgments}

\bibliographystyle{apsrev4-2}
\bibliography{main}% Produces the bibliography via BibTeX.

\end{document}

\preprint{APS/123-QED}
%\appendix
\title{\textit{Supplemental Material for}\\
Sub-Doppler cooling of a trapped ion in a phase-stable polarization gradient}

\author{Ethan Clements}
\email{eclement@mit.edu}
\author{Felix W. Knollmann}
\author{Sabrina Corsetti}
\author{Zhaoyi Li}
\author{Ashton Hattori}
\author{Milica Notaros}
\affiliation{%
Massachusetts Institute of Technology, Cambridge, Massachusetts 02139, USA
}
\author{Reuel Swint}
\affiliation{
Lincoln Laboratory, Massachusetts Institute of Technology, Lexington, Massachusetts 02420, USA
}
\author{Tal Sneh}
\affiliation{%
Massachusetts Institute of Technology, Cambridge, Massachusetts 02139, USA
}
\author{May E. Kim}
\affiliation{
Lincoln Laboratory, Massachusetts Institute of Technology, Lexington, Massachusetts 02420, USA
}
\author{Aaron D. Leu}
\affiliation{Clarendon Laboratory, Department of Physics, University of Oxford, Parks Road, Oxford OX1 3PU, United Kingdom}
\author{Patrick Callahan}
\author{Thomas Mahony}
\affiliation{
Lincoln Laboratory, Massachusetts Institute of Technology, Lexington, Massachusetts 02420, USA
}
\author{Gavin N. West}
\affiliation{%
Massachusetts Institute of Technology, Cambridge, Massachusetts 02139, USA
}
\author{Cheryl Sorace-Agaskar}
\author{Dave Kharas}
\author{Robert McConnell}
\author{Colin D. Bruzewicz}
\affiliation{
Lincoln Laboratory, Massachusetts Institute of Technology, Lexington, Massachusetts 02420, USA
}
\author{Isaac L. Chuang}
\author{Jelena Notaros}
\affiliation{%
Massachusetts Institute of Technology, Cambridge, Massachusetts 02139, USA
}
\author{John Chiaverini}\affiliation{%
Massachusetts Institute of Technology, Cambridge, Massachusetts 02139, USA
}%
\affiliation{
Lincoln Laboratory, Massachusetts Institute of Technology, Lexington, Massachusetts 02420, USA
}

\date{\today}% It is always \today, today,
             %  but any date may be explicitly specified
\maketitle

\begin{figure}[htbp]
\centering
\includegraphics[width=0.47
\textwidth]{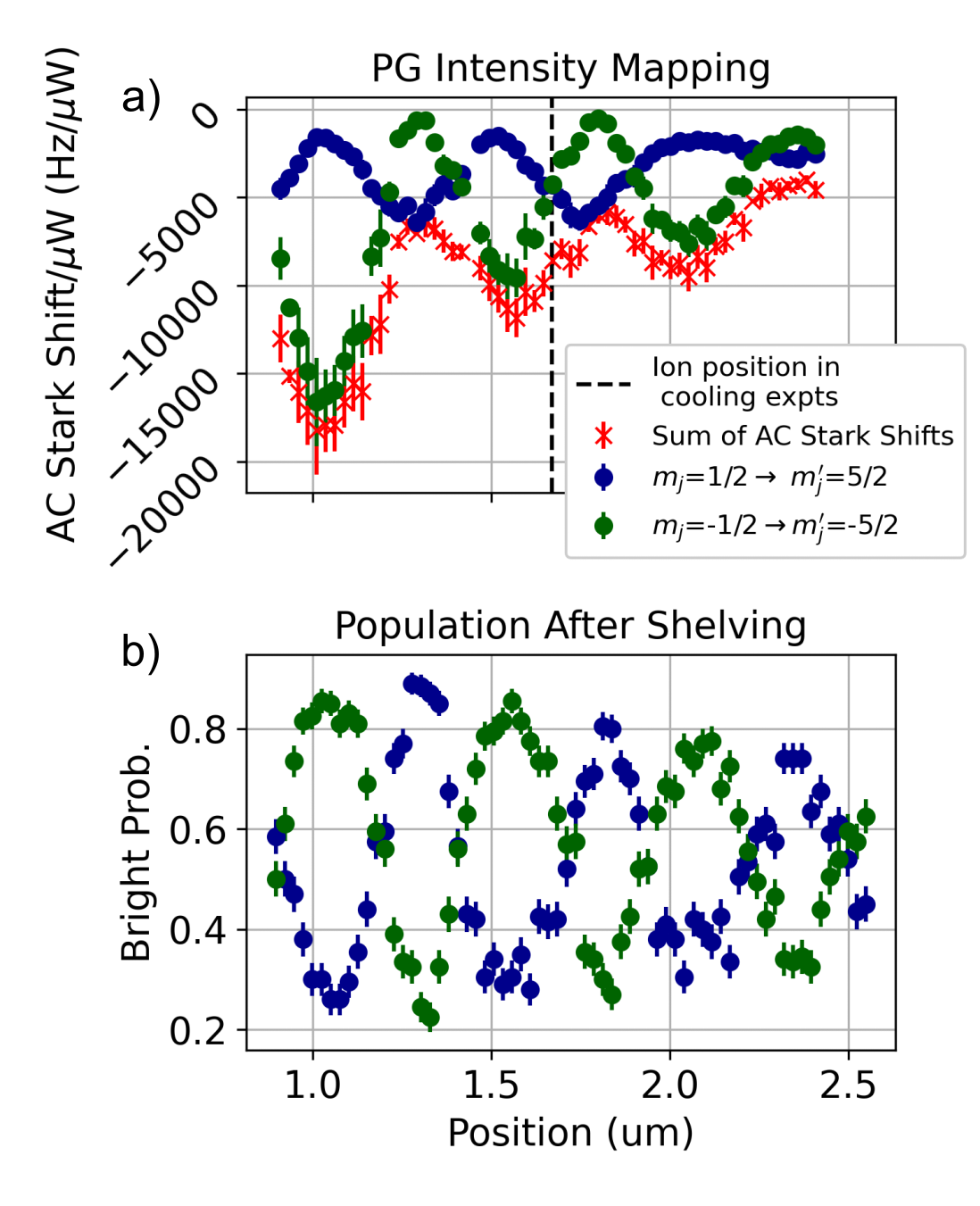}
\caption{Position and state dependence of the polarization gradient stark shift. a) We plot the light shift on the $\ket{S_{1/2}, m_{j} = 1/2}\leftrightarrow \ket{D_{5/2}, m_{j} = 5/2}$ and $\ket{S_{1/2}, m_{j} = -1/2}\leftrightarrow \ket{D_{5/2}, m_{j} = -5/2}$ transitions per $\mu$W of 422 nm light sent into the vacuum chamber as a function of the distance from the location of intended beam overlap. The sum of these shifts at each location is proportional to the local intensity. The light is detuned $2\pi\cdot160$ MHz blue of resonance. The periodic behavior on each transition shows the effect of the polarization gradient. The sum reveals the intensity variation due to the overlap of non-orthogonal polarization and H-field components and the decaying envelope of the beam profile away from the center of the overlap region. b) We provide the population after a 10 $
\mu s$ shelving pulse following Doppler cooling for ease of comparing to the optical pumping data in the main text.}
\label{fig:stark_sum}
\end{figure}
\section{Determining the optical potential landscape from Stark shifts}

We use spectroscopy of the 674 nm narrow line transition to determine the light shift of the ground states due to the polarization gradient light as shown in Fig.~S\ref{fig:stark_sum}. Cooling experiments are done at a displacement of $\sim$1.7 $\mu$m where the slope of the Stark shift on the two transitions are closest to equal and opposite. The imbalance of the intensity in regions of \(\sigma^-\) vs. \(\sigma^+\) light is caused by an intensity variation with the same period as the cycle from a \(\sigma^+\) to a \(\sigma^+\) region. This periodic fluctuation is expected due to interference from the overlap of non-orthogonal polarization components of a beam derived from free-space emission of the quasi-TE mode of light in a waveguide and the partial overlap of the H-fields of the two intersecting beams. This is discussed in more detail in ref.~\cite{corsetti_pgc}.

\begin{figure}[htbp]
\centering
\includegraphics[width=0.47
\textwidth]{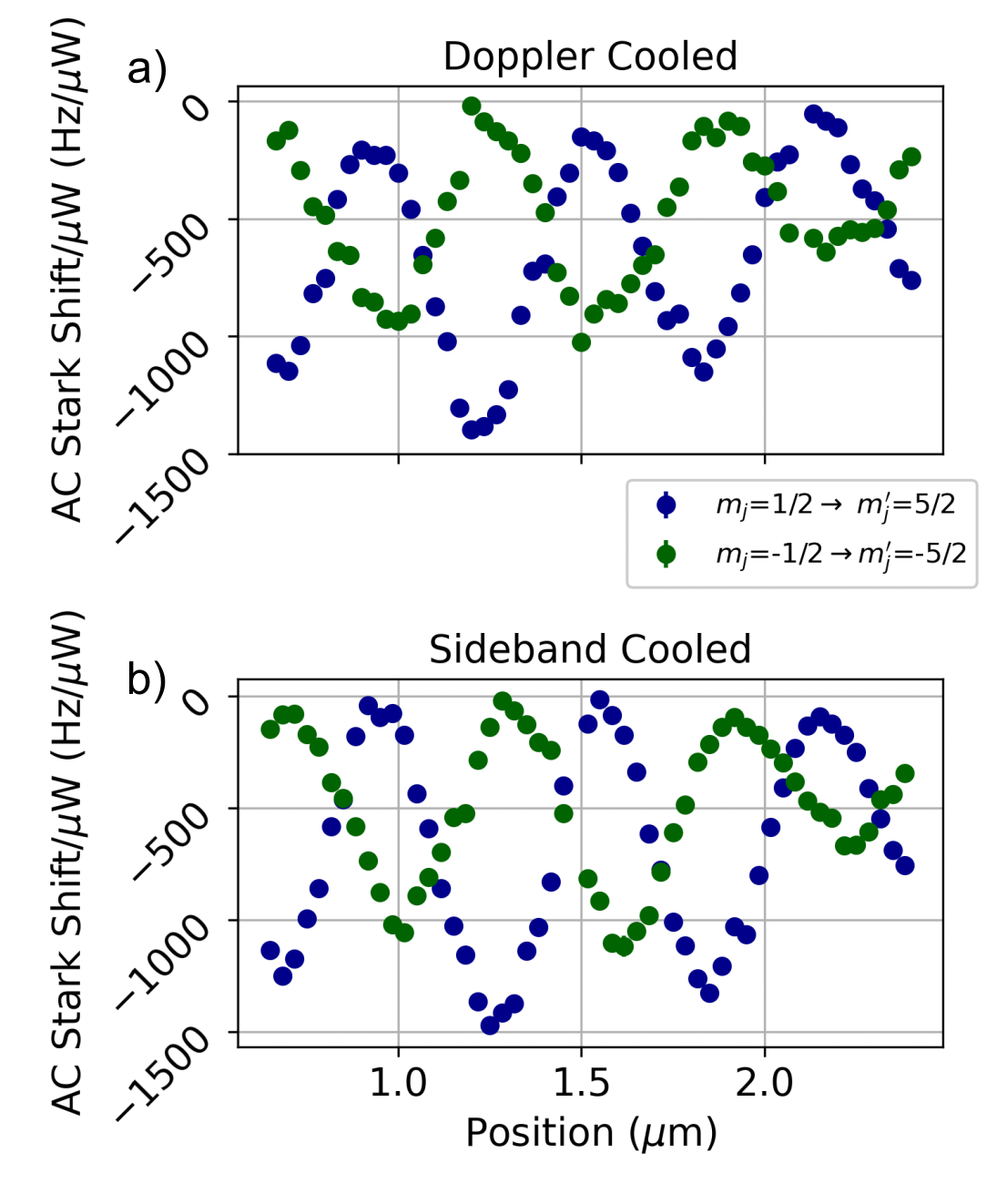}
\caption{Temperature dependence of AC-Stark shift. We plot the position dependent light shifts on the $\ket{S_{1/2}, m_{j} = 1/2}\leftrightarrow \ket{D_{5/2}, m_{j} = 5/2}$ and $\ket{S_{1/2}, m_{j} = -1/2}\leftrightarrow \ket{D_{5/2}, m_{j} = -5/2}$ transitions after Doppler cooling (a) and sideband cooling (b). This plot has lower stark shifts than others presented because data was taken on a different device with the same layout. Here the stark shift is given as Hz per $\mu$W of 422 nm light sent into the vacuum chamber. These plots demonstrate the increased contrast after cooling to \(\bar{n}\sim0.3\) vs. \(\bar{n}\sim14\). The fitted offset from full contrast is 4.8\(\pm\)1.5\% after sideband cooling and 10.2\(\pm\)2.5\% after Doppler cooling.}
\label{fig:stark_temp}
\end{figure}
\section{Analysis of the ion-motion dependence of the state-preparation contrast}
\begin{figure}[htbp]
\centering
\includegraphics[width=0.47
\textwidth]{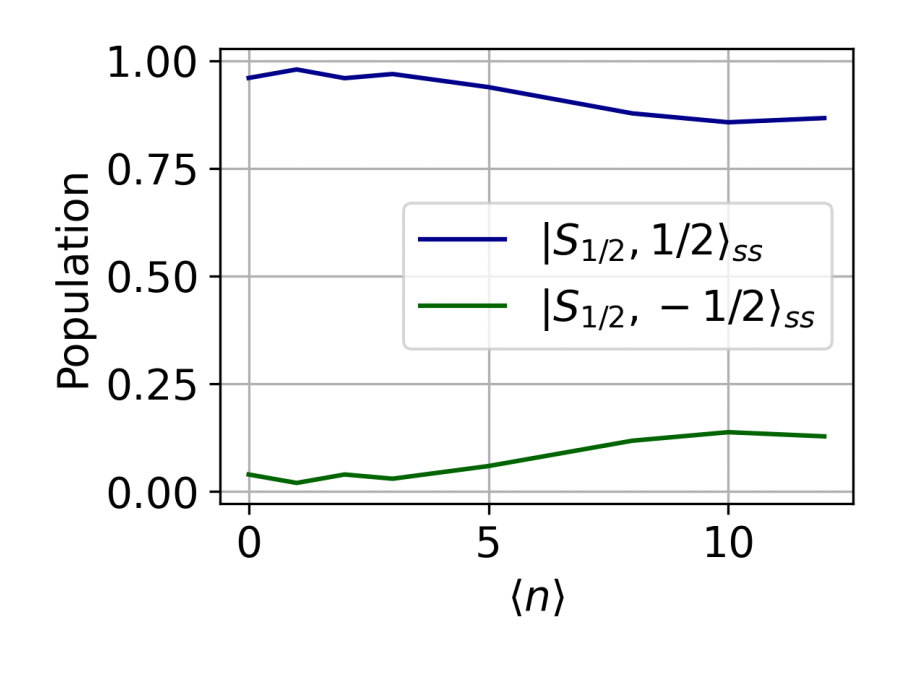}
\caption{Temperature dependence of optical pumping using structured light field. Steady state population of the simulated optical pumping sequence where the ion is placed in a location with $\sigma^+$ polarized light. The simulation is started with the ion in the $\ket{S_{1/2}, m_j = -1/2}$ ground state to avoid bias, but additional simulations were performed starting in the other ground state that yield the same results. This data shows the impact of the motional state on optical pumping and suggests that only coherent states of the ion motion will yield a location where optical pumping is maximized. The resulting populations show good agreement with experimental data.}
\label{fig:prep_motion_dep}
\end{figure}
The phase of the polarization gradient sampled by the ion depends on the ion motional state since the ion wavepacket size is related to the square root of the average motional excitation measured in quanta of the motional mode energy. This can be seen in the equation for the coupling operator which describes the dipole interaction of the polarization gradient:
\begin{align}
    \nonumber\widehat{G}^+ &= \widehat{\mathbf{d}}^+ \cdot \mathbfcal{E} \\
    \nonumber&=  \Omega \left( \frac{\widehat{L}_+}{\sqrt{2}} \left( \cos(k\widehat{z} + \phi) + \sin(k\widehat{z} + \phi) \right) \right. \\
    &\left. + \frac{\widehat{L}_-}{\sqrt{2}}\left( -\cos(k\widehat{z} + \phi) + \sin(k\widehat{z} + \phi) \right) \right) 
\end{align}

(This equation is further detailed later in the supplement). The effect of the motion arises from the $k\widehat{z}$ term which can be rewritten in the Lamb-Dicke regime as $\eta ( a^{\dag}\!+\!a)$. As a result there is now a Fock state dependent phase in the sine and cosine terms which means there is no unique phase for a thermal state that will result in the coupling operator only having a $\widehat{L}_+$ or $\widehat{L}_-$ term. This can also be viewed as an ion in a thermal state sampling various phases of the polarization gradient due to the spread of the ion wavefunction. This effect smears out the contrast of the state preparation fringes as shown in Figs.~$1$ and~S\ref{fig:stark_temp}. These data were taken after Doppler cooling to an average \(\langle n\rangle\) of \(\sim\)14. The \(\sim\)70\% contrast observed is mainly limited by this effect. We expect an additional \(\sim\)1-2 quanta of heating from photon scattering during the optical pumping. We also model the population dynamics in the low saturation limit, at a phase that gives $\sigma^+$ polarization, starting with different Fock states to show the impact the motional state has on state preparation. These results are shown in Fig~S\ref{fig:prep_motion_dep} where the simulation begins with all of the population in the $\ket{S_{1/2}, m_j = -1/2}$ state.

Our model of the projection of the quasi-TE waveguide mode into free space suggests that the \(\pi\)-polarized component of light would be negligible at the correct overlap position of the two beams. However, in our system we are operating on the edge of the beam which leads to a simulated upper bound on the fraction of \(\pi\)-polarized light of 8-12\% (meaning \(\ge\)88\% purity of \(\sigma^{+/-}\)). Fitting the oscillations for the sideband cooled case in Fig.~S\ref{fig:stark_sum} we see a 4.8\(\pm\)1.5\% offset from perfect extinction. Due to the Clebsch Gordon factor for \(\sigma^{+/-}\) being twice that for the \(\pi\) transition, this allows us to experimentally bound the fraction of \(\pi\)-polarized light to 7-13\%, in agreement with the model. Beyond the effect of \(\pi\)-polarized light, the decay in contrast can be explained by imperfect balancing of the intensities from the two beams away from the center of the intersection causing an elliptical polarization, residual excitation of the motion, and misalignment of the quantization magnetic field from the ideal direction perpendicular to the plane of the trap. Both of the last effects were independently measured to be negligible.

\section{Stabilization of the ion position in a polarization gradient}

By measuring ion position drifts after varying lengths of photoionization light illumination (405 nm and 461 nm), we determine that relative stability between the ion and polarization gradient is primarily limited by drifts of the ion location arising from charging of the trap surface during ion loading. Sending light through the grating emitters had no effect on the ion position across varying intensities and exposure durations. Following a minute of photoionization light exposure, we measure a discrete jump of 50-100 nm and then a \(<\)10 nm/hour ion drift rate with an exponential relaxation time constant of a couple hours. Further characterization of the stability is presented in Ref~\cite{corsetti_pgc}. To minimize the impact of these drifts we utilize the previously mentioned position dependent state preparation outcome [Fig.~$1$(c)] as an error signal and feed back on the axial ion position by changing the DC voltages of our surface electrode ion trap. This nulls the relative drift as determined from the closed loop measurements. Each loop of feedback takes a few seconds and is performed every 2 to 3 minutes. Ultimately this instability could be passively reduced by loading further away from the polarization gradient location or reducing the photo-ionization laser intensity at the trap surface. By implementing this low-duty-cycle, closed-loop feedback on ion position we are able to maintain the necessary long-term stability to demonstrate position dependent cooling in a phase-stable polarization gradient.  

\section{AOM efficiency at different detunings}
\begin{figure}[htbp]
\centering
\includegraphics[width=0.45\textwidth]{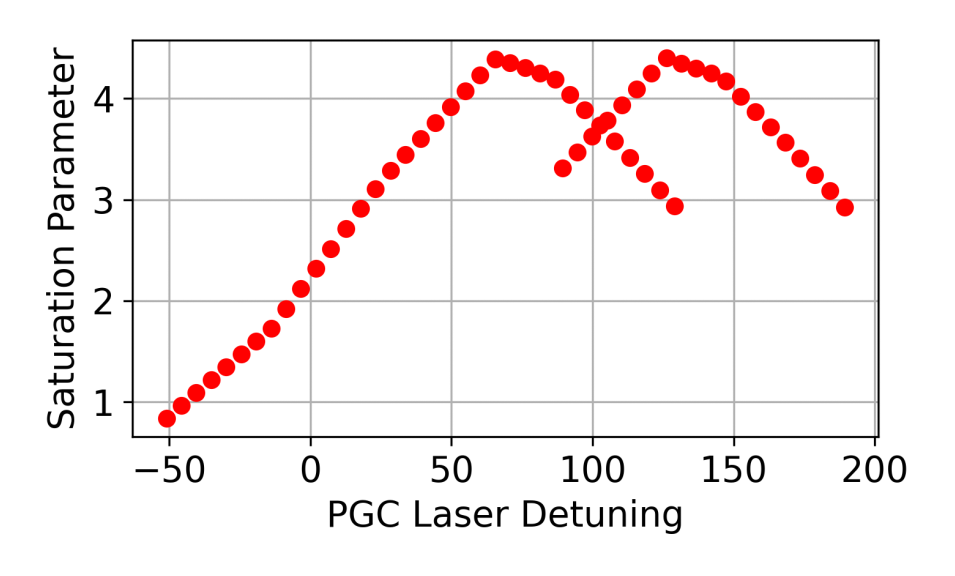}
\caption{AOM diffraction efficiency. This plot of the saturation parameter as a function of detuning shows the effect of our AOM diffraction bandwidth. The second set of points from $2\pi\cdot80$ to $2\pi\cdot190$ MHz is obtained when locking the 422 nm laser to a second peak in the Rb saturated absorption spectrum $2\pi\cdot60$ MHz red of the original peak.}
\label{fig:aom}
\end{figure}

Due to the limited overlap of the two beams at the trapping axis of the ion as displayed in Fig.~$1$(b) and 10 dB excess loss at the fiber to waveguide interface totaling almost 50 dB of attenuation from light sent into the vacuum chamber to light at the ion, there was not enough laser power overhead to maintain the desired intensity across the full range of detunings we tested. We include a plot of the resonant saturation parameter as a function of detuning from atomic resonance due to the limited bandwidth of the AOM we used to scan the 422 nm laser frequency (Fig.~S\ref{fig:aom}).

\section{Modeling of polarization gradient cooling}
In this section we detail the theoretical model used with QuTip to simulate our system and detail the semi-classical model as a comparison. We describe the simplified laser configuration and the relevant couplings for a system that only considers the transitions between the $S_{1/2}$ and $P_{1/2}$ sub-spaces and extend this to a system which considers transitions between the $S_{1/2}$, $P_{1/2}$, and $D_{3/2}$ sub-spaces to determine the impact state coherences have on cooling. We also detail how the models can be extended to a running-wave polarization gradient configuration and compare the cooling dynamics for this configuration against the phase-stable polarization gradient. Included is a write-up of a semi-classical approach to modeling this system which utilizes the work presented in Ref.~\cite{cirac1993laser}. This serves as a good representation over a fixed parameter range and provides intuition for these cooling processes but is limited in broad application.

\paragraph{Laser Configuration}\mbox{}\\
We consider two counter-propagating waves along the \( z \)-axis with orthogonal polarizations. The polarization vectors are expressed as:
\begin{equation}
\boldsymbol{\epsilon}_{\mathrel{\mathop{\rightleftarrows}}} = \frac{1}{\sqrt{2}} (\wideparen{\mathbf{x}} \pm \wideparen{\mathbf{y}}),
\end{equation}
where the arrows denote the respective propagation directions of the waves. Here, \( \wideparen{\mathbf{x}} \) and \( \wideparen{\mathbf{y}} \) represent unit vectors along the \( x \) and \( y \)-axes, respectively. This creates a phase-stable polarization gradient with polarizations varying between circular ($\sigma^\pm$) and linear (lin) at different $z$ coordinates or phases. For example, at $z=0$, $\phi=\pm\pi/4+n\pi$ corresponds to $\sigma^\pm$-polarized light, and when $\phi=n\pi/2$, the light is linearly polarized. The light field can therefore be described as:
\begin{equation}
\widehat{\mathbf{E}} = \widehat{\mathbfcal{E}} e^{-i\omega t} + \text{h.c.},
\end{equation}
with the time-independent part given by:
\begin{equation}
\widehat{\mathbfcal{E}}(\widehat{\mathbf{R}}) = \mathcal{E}_0 \left( \boldsymbol{\epsilon}_\rightarrow e^{i(\phi + \mathbf{k}_\rightarrow\cdot\widehat{\mathbf{R}}))} + \boldsymbol{\epsilon}_\leftarrow e^{-i(\phi + \mathbf{k}_\leftarrow\cdot\widehat{\mathbf{R}}))} \right), 
\end{equation} where $\widehat{\cdot}$ denotes operators and \( k_{\rightleftarrows}\) are the wave vectors of the two counter propagating beams.

\paragraph{Hamiltonian for $S_{1/2}$ and $P_{1/2}$ subspace }\mbox{}\\
The full system Hamiltonian is given by:
\begin{equation}
\widehat{H} = \widehat{H}_0 + \widehat{H}_{\mathrm{trap}} + \widehat{H}_{\mathrm{int}},
\end{equation}
where \( \widehat{H}_0 \) is the free internal Hamiltonian associated with the atomic transition frequency \( \omega_0 \). The harmonic trapping Hamiltonian is expressed as:
\begin{equation}
\widehat{H}_{\mathrm{trap}} = \hbar\omega \left( \widehat{N} + \frac{1}{2} \right),
\end{equation}
where \( \omega \) represents the trap frequency and \( \widehat{N} \) is the number operator; here we focus on a single vibrational mode but maintain generality to multiple modes. The light-atom interaction term is:
\begin{equation}
\widehat{H}_{\mathrm{int}} = -\widehat{\mathbf{d}} \cdot \widehat{\mathbf{E}}(\widehat{\mathbf{R}}, t),\vspace{0.4cm}
\end{equation}
where \( \widehat{\mathbf{d}} \) is the dipole operator and \( \widehat{\mathbf{E}}(\widehat{\mathbf{R}}, t) \) is the electric field at the atomic position \( \widehat{\mathbf{R}} \). 
For convenience, we define the \textit{excitation dipole operator} as:
\begin{equation}
\widehat{\mathbf{d}}^+ = \widehat{\Pi}_e \widehat{\mathbf{d}} \widehat{\Pi}_g,
\end{equation} where \(\widehat{\Pi}_e\) and \(\widehat{\Pi}_g\) are the projection operators onto the excited \( P_{\frac{1}{2}} \) and ground \( S_{\frac{1}{2}} \) subspaces, respectively. Accordingly, we introduce the {\it excitation (relaxation) coupling operators} \(\widehat{G}^+\) and \(\widehat{G}^-\), with the relation \(\widehat{G}^+ = (\widehat{G}^-)^\dagger\). These operators are defined by:
\begin{equation}
\widehat{G}^+ = \widehat{\mathbf{d}}^+ \cdot \widehat{\boldsymbol{\mathcal{E}}}.
\end{equation}
Switching to the laser frame and performing the rotating wave approximation (RWA) \cite{fujii2013introduction}, we get:
\begin{equation}
\widetilde{H_0+H_{\mathrm{int}}} = \begin{bmatrix}
  -\frac{\delta}{2} &  -\widehat{G}^+\vspace{0.2cm}\\
     -\widehat{G}^- & \frac{\delta}{2}
\end{bmatrix}, 
\end{equation}
where $\delta=\omega-\omega_0$ is the laser detuning. The matrix is organized into blocks corresponding to the subspace projectors \(\widehat{\Pi}_e\) and \(\widehat{\Pi}_g\), with identity operators omitted. Consequently, the overall Hamiltonian (up to an overall shift by \(\delta/2\)) is given by:

\begin{equation}
\widetilde{H} = \widehat{H}_\mathrm{trap} 
-\delta \widehat{\Pi}_e - \widehat{G}, \quad \widehat{G} = \widehat{G}^+ + \widehat{G}^-.
\end{equation}

To study the effects of spontaneous emission, we define the transition operators:
\begin{equation}
\widehat{L}_q = \boldsymbol{\epsilon}_q \cdot \wideparen{\mathbf{r}}^-, 
\end{equation}
for \( q \in \{-, 0, +\} \). The polarization vectors \(\boldsymbol{\epsilon}_q\) are defined with respect to the Cartesian coordinates as:
\begin{equation}
\begin{aligned}
    \boldsymbol{\epsilon}_\pm &= \mp\frac{1}{\sqrt{2}} (\wideparen{\mathbf{x}} \pm i\wideparen{\mathbf{y}}), \\
    \boldsymbol{\epsilon}_0 &= \wideparen{\mathbf{z}},
\end{aligned}
\end{equation}
and \(\wideparen{\mathbf{r}} = \frac{\mathbf{r}}{r}\) is the unit radial vector which lies in the xy plane orthogonal to the quantization axis, and \(\wideparen{\mathbf{r}}^- = \widehat{\Pi}_g \wideparen{\mathbf{r}} \, \widehat{\Pi}_e\) is the corresponding relaxation operator. Note here the transition operators satisfy the completeness relation:
\begin{equation}
\sum_q\widehat{L}_q^\dagger \widehat{L}_q = \widehat{\Pi}_e.
\end{equation}

We define the Rabi frequency using the following relation:
\begin{equation}
\langle S_{\pm\frac{1}{2}} | \widehat{\mathbf{d}}^+ \cdot \boldsymbol{\epsilon}_+ | P_{\mp\frac{1}{2}} \rangle = \mp\sqrt{\frac{2}{3}}\frac{\Omega \hbar}{\mathcal{E}_0}.
\end{equation}
Note that in our case $\Omega$ is real. Now we explicitly evaluate:
\begin{widetext}

\begin{align}
    \nonumber\widehat{G}^+ &= \widehat{\mathbf{d}}^+ \cdot \mathbfcal{E} \\
    \nonumber&=  \Omega \left( \frac{\widehat{L}_+}{\sqrt{2}} \left( \cos(k\widehat{z} + \phi) + \sin(k\widehat{z} + \phi) \right) + \frac{\widehat{L}_-}{\sqrt{2}}\left( -\cos(k\widehat{z} + \phi) + \sin(k\widehat{z} + \phi) \right) \right) \\
    &= \begin{bmatrix}
        0 & 0 & 0 & \frac{\Omega \left( \cos(k\widehat{z} + \phi) + \sin(k\widehat{z} + \phi) \right)}{\sqrt{3}} \\
        0 & 0 & \frac{\Omega \left( -\cos(k\widehat{z} + \phi) + \sin(k\widehat{z} + \phi) \right)}{\sqrt{3}} & 0 \\
        0 & 0 & 0 & 0 \\
        0 & 0 & 0 & 0
    \end{bmatrix},
\end{align}
where the basis of this operator is spanned by the $S_{1/2}$ and $P_{1/2}$ Zeeman state vectors.
\end{widetext}

\paragraph{Hamiltonian including the $S_{1/2}$,$P_{1/2}$, and $D_{3/2}$ sub-spaces}\mbox{}\\
To model the impact of coherences in our system between the $S_{1/2}$ and $D_{3/2}$ Zeeman states we further expand the Hamiltonian to include terms for the $D_{3/2}$ Zeeman states and a term to account for the Zeeman effect when there is a magnetic field applied. The term describing the Zeeman shift from an applied quantization field is given by:
\begin{equation}
    \begin{aligned}
        H_{Z} = \mu_{B}B \sum_m\sum_l m g_{l}\ket{l,m}\bra{l,m} 
    \end{aligned}
\end{equation}

\noindent where $B$ is the magnitude of the magnetic field, $\mu_B$ is the Bohr magneton, $l$ is the electronic state, $m$ is the quantum number describing the Zeeman sub-levels, and $g_l$ is the Land\'e g-factor for each state. Transition operators would be constructed identically for the addition of the metastable state $D_{3/2}$. Coupling between levels not caused by the polarization gradient, here denoted as $\widehat{\mathcal{G}}^+$, would be simply described as:
\begin{equation}
    \begin{aligned}
        \widehat{\mathcal{G}}^+ = \sum_q c_{q}\Omega_{q}\widehat{L}_q
    \end{aligned}
\end{equation}
where $c_q$ is the transition specific Clebsch-Gordon coefficent, and $\Omega_q$ is the transition specific Rabi frequency. If effects of polarization impurity or imbalance are included in the modeling this could effect the relative values of the transition specific Rabi frequencies and if deviations in the TE components of the polarization gradient are included $\mathcal{G}$ terms for the $q=0$ component of the polarization would also be included. 

For the near resonant case of the repumper and polarization gradient light one can similarly take the rotating wave approximation for this three electronic state system~\cite{rei1996cooling} and obtain the Hamiltonian for the internal state basis and the light-matter interaction: 

\begin{equation}
\widetilde{H_0+H_{\mathrm{int}}} = \begin{bmatrix}
  0 &  -\widehat{G}^+ & -\widehat{\mathcal{G}}^+\vspace{0.2cm} \\
     -\widehat{G}^- & \delta_{ps} & 0 \vspace{0.2cm} \\
     -\widehat{\mathcal{G}}^- & 0 & \delta_{pd}
\end{bmatrix}
\end{equation}

\noindent where \(\delta_{ps} \) and \(\delta_{pd} \) are the detuning from resonance for the 422 nm PGC laser and the 1092 nm repumper laser respectively. This then results in an overall Hamiltonian of:

\begin{align}
\widetilde{H} = \widehat{H}_\mathrm{trap} 
+ \delta_{ps} \widehat{\Pi}_s + \delta_{pd} \widehat{\Pi}_d - \widehat{G} -\widehat{\mathcal{G}}, \quad \nonumber \\
\widehat{G} = \widehat{G}^+ + \widehat{G}^-, \nonumber\\
\widehat{\mathcal{G}} = \widehat{\mathcal{G}}^+ + \widehat{\mathcal{G}}^-
\end{align}
With this we now have the necessary Hamiltonian to describe the system for both considerations of the electronic state space. To describe the dissipative dynamics, we now need to derive the collapse operators for our master equation.
\begin{widetext}
\paragraph{Master Equation}\mbox{}\\
Now, consider the dynamics of the system as described by the master equation with density matrix \(\rho\):
\begin{equation}
    \dot{\rho}=-i\left[H,\rho\right]+\frac{\Gamma}{2}\sum_{\boldsymbol{\epsilon}\perp\mathbf{k}}\int d^2\,\Omega(\mathbf{k})\left(2\widehat{C_{\mathbf{k},\boldsymbol{\epsilon}}}\rho\widehat{C_{\mathbf{k},\boldsymbol{\epsilon}}}^\dag-\left\{\widehat{\Pi_e},\rho\right\}\right)\label{equ:master}
\end{equation}
Here, $\widehat{C_{\mathbf{k},\boldsymbol{\epsilon}}}$, the jump operators, describe the spontaneous decay and recoil processes acting on the internal and motional registers, respectively, given a transition linewidth \(\Gamma\). These operators are defined as:
\begin{equation}\widehat{C_{\mathbf{k},\boldsymbol{\epsilon}}}=\sqrt{\frac{3}{8\pi}}e^{-i\mathbf{k}\cdot\widehat{\mathbf{R}}}\widehat{\wideparen{\mathbf{r}}}^{-}\cdot\boldsymbol{\epsilon},\end{equation}
Here the prefactor can be determined by normalization of the completeness condition
\begin{equation}\sum_{\boldsymbol{\epsilon}\perp\mathbf{k}}\int d^2\,\Omega(\mathbf{k})\widehat{C_{\mathbf{k},\boldsymbol{\epsilon}}}^\dag\widehat{C_{\mathbf{k},\boldsymbol{\epsilon}}}=\widehat{\Pi}_e,\end{equation}
where we have accounted for all possible recoil directions by integrating over the solid angle $\Omega(\mathbf{k})$ for the spontaneous emission of a photon with wavevector $\mathbf{k}$ in the laser frame. Additionally, we sum over the two allowed polarization states of the photon that are perpendicular to $\mathbf{k}$.

\begin{figure*}[htbp!]
\centering
\includegraphics[width=.8
\textwidth]{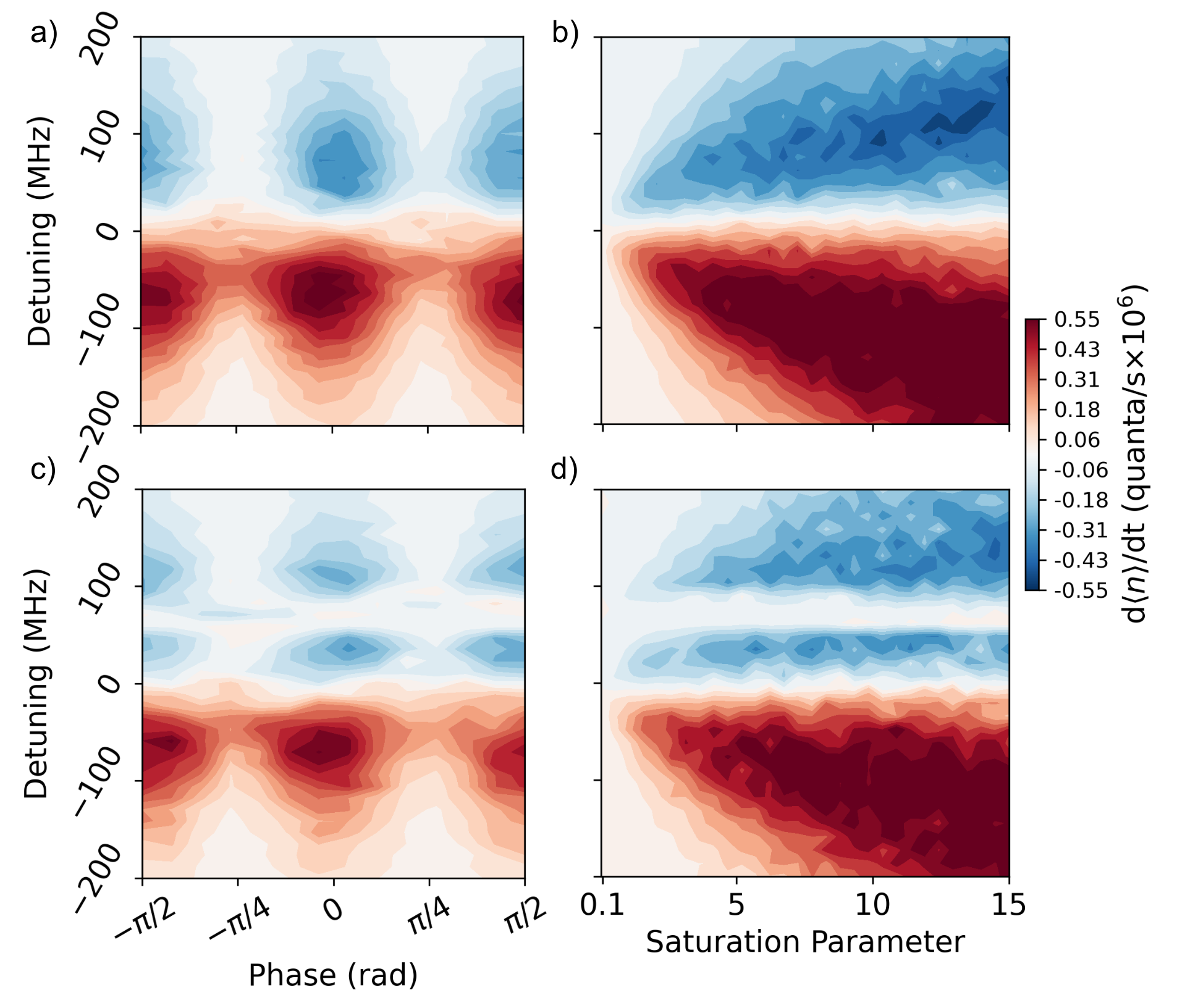}
\caption{Numerical simulations of polarization gradient cooling rate. Our numerical simulations show qualitative agreement with the corresponding experimental results in the main text. We examine the cooling/heating rate (\(d\langle n \rangle/dt \)) as a function of polarization gradient phase, detuning, and saturation parameter to show the characteristic phase- and detuning-dependent cooling of PGC.}
\label{fig:model_overview}
\end{figure*}

When studying 1D cooling dynamics we improve the speed of the simulation by simplifying the jump operators, which include the spontaneous decay and photon recoil dynamics, by using a discretized probability distribution instead of a continuous probability distribution, following previous works~\cite{joshi2020polarization,Molmer93}. we can further simplify Eq. \ref{equ:master} by manipulating the jump operator term:

\begin{align}
\label{equ:jump_1d}
    &\nonumber\int \sum_{\boldsymbol{\epsilon}\perp\mathbf{k}} d^2\,\Omega(\mathbf{k})\widehat{C_{\mathbf{k},\boldsymbol{\epsilon}}}\rho\widehat{C_{\mathbf{k},\boldsymbol{\epsilon}}}^\dag\\
    =&\sum_{q\in\{-,0,+\}}\int d^2\,\Omega(\mathbf{k})\widehat{C_{k_z,\boldsymbol{\epsilon}_q}}\rho\widehat{C_{k_z,\boldsymbol{\epsilon}_q}}^\dag-\int d^2\,\Omega(\mathbf{k})\customwidehat{C_{k_z,\wideparen{\mathbf{k}}}}\rho\customwidehat{C_{k_z,\wideparen{\mathbf{k}}}}^\dag,
\end{align}
\noindent where in the second term  of the equation we correct for the over counting of the $\mathbf{k}$ projection onto the 1-D cooling axis. In Eqn.~\ref{equ:jump_1d} we have defined our one-dimensional $\widehat{C_{k_z,\epsilon}}$ operators only in the $z$ (cooling) direction by tracing off the $x$ and $y$ registers  :
\begin{equation}
\widehat{C_{k_z,\epsilon}}=\sqrt{\frac{3}{8\pi}}e^{-ik_z\widehat{Z}}\widehat{\wideparen{\mathbf{r}}}^{-}\cdot\boldsymbol{\epsilon}.
\end{equation}

\noindent Expanding the second term in cylindrical coordinates, there is $d^2\Omega(\mathbf{k}) =\dfrac{dk_z}{k}dk_\phi$, where $k_z$, $k_\phi$ are the components of $\mathbf{k}$ in cylindrical coordinates, and $k$ is the magnitude of $\mathbf{k}$ which is fixed. We get
\begin{align}
    \sum_{q\in\{-,0,+\}}\int_{-k}^{k}\frac{dk_z}{k}\frac{3}{4}\frac{|k_q|^2}{k^2}e^{-ik_z\widehat{Z}}\widehat{L_q}\rho\widehat{L_q}^\dag e^{ik_z\widehat{Z}},
\end{align}
where $k_q$ is $\mathbf{k}$ in polarization coordinates, i.e. $\mathbf{k}\cdot\boldsymbol{\epsilon}_q$. Therefore, $\frac{k_\pm^2}{k^2}=\frac{1}{2}\left(1-\frac{k_z^2}{k^2}\right)$, and the total jump term is given by:
\begin{align}
    \sum_{q\in\{-,0,+\}}\int_{-k}^{k}\frac{dk_z}{k}\frac{3}{4}\left(1-\frac{|k_q|^2}{k^2}\right)e^{-ik_z\widehat{Z}}\widehat{L_q}\rho\widehat{L_q}^\dag e^{ik_z\widehat{Z}}.
\end{align}
For easier sampling, previous studies have substituted the continuous distribution with a discretized probability supported on $k_z \in \{\pm k, 0\}$: 
\begin{align}
    \sum_{q\in\{-,0,+\}}\sum_{k_z\in\{-k,0,+k\}}N_{k_z,q}e^{-ik_z\widehat{Z}}\widehat{L_q}\rho\widehat{L_q}^\dag e^{ik_z\widehat{Z}}.
\end{align}
To match the first two moments of the distribution, we will define \(N_{k,0} = \frac{1}{10}\), \(N_{0,0} = \frac{8}{10}\), \(N_{-k,0} = \frac{1}{10}\), and \(N_{k,\pm} = \frac{1}{5}\), \(N_{0,\pm} = \frac{3}{5}\), \(N_{-k,\pm} = \frac{1}{5}\). The simplified overall dynamics are given by:

\begin{equation}
    \dot{\rho}=-i\left[\widehat{H},\rho\right]+\Gamma\sum_{\substack{k_z\in\{-k,0,+k\} \\ q\in\{-,0,+\}}}N_{k_z,q}e^{-ik_z\widehat{Z}}\widehat{L_q}\rho\widehat{L_q}^\dag e^{ik_z\widehat{Z}}-\frac{\Gamma}{2}\left\{\widehat{\Pi_e},\rho\right\}\label{equ:discretized}.
\end{equation}
\end{widetext}
For the Qutip simulations the collapse operators are entered as:
\begin{equation}
    \widehat{C_{\mathbf{k},\boldsymbol{\epsilon_q}}} =\sum_{\substack{k_z\in\{-k,0,+k\} \\ q\in\{-,0,+\}}}\sqrt{\Gamma}\sqrt{N_{k_z,q}}e^{-ik_z\widehat{Z}}\widehat{L_q}
\end{equation}
In the case of the multi-electronic state simulations the decay rate was multiplied by the branching ratio for the different decay pathways and the Lamb-Dicke parameter was scaled for the given transition wavelength.

\begin{figure*}[htbp]
\centering
\includegraphics[width=1.0\textwidth]{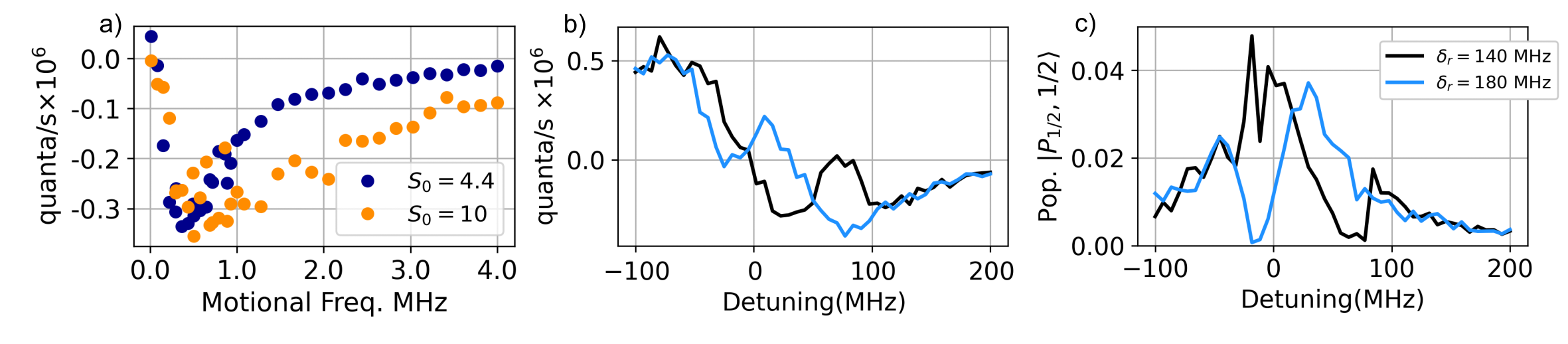}
\caption{Simulated effect of motional frequency and dark state location on cooling. Here we provide additional simulations of the cooling rate dependence on the motional frequency and how the dark resonance location varies in the cooling rate plot and fluorescence spectrum as the repumper detuning $\delta_r$ is varied. In all of these plots $\phi = 0$, the saturation parameter for the PGC light = 4, and the saturation parameter for the repumper is 5000. In (a) the PGC detuning was set to $2\pi\cdot140$~MHz. Here we see that the cooling rate falls off quickly at motional frequencies higher than $2\pi\cdot1$~MHZ. This range broadens to larger motional frequencies for higher saturation parameters. In (b) and (c) the detuning of the repumper is varied and shows how the placement of dark resonance can greatly impact the effectiveness of polarization gradient cooling.}
\label{fig:dark_state}
\end{figure*}

\begin{figure*}[htbp]
\centering
\includegraphics[width=.95\textwidth]{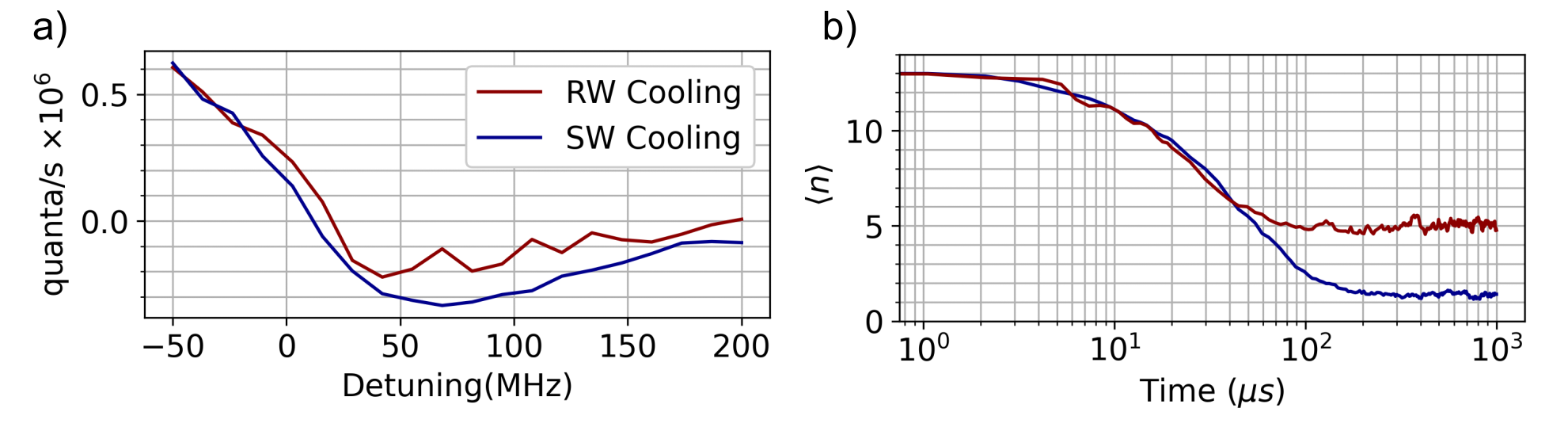}
\caption{ Comparison of running wave (RW) and standing wave (SW) polarization gradient cooling. In this comparison a detuning for the running wave of 100 kHz was chosen and simulations were again performed with a PG saturation parameter of 4 and $\phi = 0$. In (a) we can see the reduction in both heating and cooling as a function of detuning when compared against the phase stable version. This is due to averaging over the phase dependence of the polarization at each detuning. In (b) we simulated a time series of cooling with the two methods where we see a steady state $\langle n \rangle$ of $\sim 5$ for running wave cooling and $\sim 1.5$ for phase stable cooling for global laser detuning of $2\pi\cdot140$~MHZ.}
\label{fig:rw_v_sw}
\end{figure*}

\paragraph{Simulations of Hamiltonians}\mbox{}\\
Using the described Hamiltonians and collapse operators we can generate simulations of the system dynamics. We plot all simulated parameter range scans for both the model which only includes the $S_{1/2}$ and $P_{1/2}$ Zeeman manifolds and the model which has the $D_{3/2}$ manifold added. Plots of the frequency, amplitude, and phase dependence are shown in Fig.~S\ref{fig:model_overview}. A larger parameter range than what was experimentally explored is shown to better illustrate the heating effects at negative laser detuning. We observe similar cooling rates across plots but observe the presence of a region of elevated heating at $\approx 2\pi\cdot75$ MHz in the three electronic state system corresponding to coherences between the $S_{1/2}$ and $D_{3/2}$ states. This simulated data assumes a 1092 nm (repumper for connecting $D_{3/2}$ to $P_{1/2}$) saturation parameter of 5000 determined from measurements of laser power and beam waist radius and a detuning of $2\pi\cdot140$ MHz which was determined by matching the location of the dark resonance in measured ion fluorescence data. In Fig.~S\ref{fig:dark_state} (a) we plot the simulated cooling rate at a detuning of $2\pi\cdot140$ MHz and saturation parameter of 4 as the motional frequency of the ion is varied. This shows how the cooling rate varies at different motional frequencies. To achieve comparable cooling rates at higher frequencies larger saturation intensities will be needed. In Fig.~S\ref{fig:dark_state}(b) and (c) we display the effect of changing the repumper laser detuning and saturation parameter has on the steady state population of the $P_{1/2}$ states and the cooling rate. In all simulations presented here we are using a magnetic field of 2.55 G aligned perpendicular to the linear polarization directions. This is the field amplitude measured in our experiment. This plot illustrates how the dark resonances can vary depending on experimental parameters.

While this is not the direct focus of this work we also implemented a QuTip simulation of cooling with a running wave polarization gradient to verify the claims that cooling with a running wave (RW) increases the steady state temperature by a factor of two. This simulation is implemented by modifying the code for simulating cooling only considering the $S_{1/2}$ and $P_{1/2}$ Zeeman manifolds. The running wave is incorporated by adding time dependence to the Hamiltonian which can simply be added by modifying the light field to include a relative laser detuning between the two beams of $\Delta_{\rm rw}$. This detuning is equally split between the two beams comprising the polarization gradient. This can be written as:
\begin{align}
\widehat{\mathbf{E}} = \mathcal{E}_0 \left( \boldsymbol{\epsilon}_\rightarrow e^{-i(\omega - \Delta_{\rm rw}/2) t} e^{i(\phi + \mathbf{k}_\rightarrow\cdot\widehat{\mathbf{R}}))} + \right.\notag \\
\left. \boldsymbol{\epsilon}_\leftarrow e^{-i(\omega + \Delta_{\rm rw}/2) t}e^{-i(\phi + \mathbf{k}_\leftarrow\cdot\widehat{\mathbf{R}}))} \right)  + \text{h.c.},
\end{align}
After the rotating wave approximation is applied this time dependent phase will remain in the electric field. The time dependence can be easily separated from the other phase terms and can be included in the QuTip simulation as
\begin{equation}
e^{\pm i\Delta_{\rm rw}t/2}
\end{equation}
for each electric field. Alternatively, the time dependence can be included in the existing formulation of the coupling operator as an additional phase term of $\Delta_{\rm rw}t$ in the sine and cosine terms. Results showing the simulated frequency dependence of the cooling and time dynamics of the average excitation in the Fock basis can be seen in Fig.~S\ref{fig:rw_v_sw}. These simulations use the same saturation parameter of 4, motional frequency of $2\pi\cdot1$ MHz and a relative beam detuning of 100 kHz for the running wave polarization gradient. In Fig.~S\ref{fig:rw_v_sw} a) we see that the cooling rate is reduced but the relative frequency dependence remains. However, in Fig.~S\ref{fig:rw_v_sw}
b) we see that the steady state temperature for RW cooling is more than a factor of 2 larger. This is likely a result of not searching for the optimum RW detuning or looking for and comparing the exact parameters where both cooling methods result in their optimum steady state temperatures. This does however demonstrate the difference in the achievable temperature limits and the higher value for running wave cooling. In future work one could explore running wave polarization gradient cooling using integrated photonics and perform a more direct theoretical and experimental comparison of the two cooling methods.

\begin{widetext}
\paragraph{Semiclassical Theory}
In the semiclassical theory, we approximate the motional dynamics by a classical process and begin with considering the $S_{1/2}$ and $P_{1/2}$ manifolds as a two-level system (ground and exited states denoted $g$ and $e$, respectively). Therefore, for the Hamiltonian, we only consider the internal DOFs for the density matrix $\rho^\prime=\mathrm{tr}_{\mathrm{ext}}(\rho)$, i.e., $H^\prime = H_0 + H_\mathrm{int}$, and we replace the position operators $\widehat{z}$ with a classical variable. 

\begin{equation}
    \dot{\rho^\prime}=-i\left[\widehat{H}^\prime,\rho^\prime\right]+\frac{\Gamma}{2}\left(2\widehat{L_q}\rho^\prime\widehat{L_q}^\dag-\left\{\widehat{\Pi}_e,\rho^\prime\right\}\right)\label{equ:traced}
\end{equation}
In block matrix form, this is 
\begin{equation}
\label{equ:effective_gs_dynamics}
\partial_t \begin{bmatrix}
    \rho_{ee} & \rho_{eg} \\
    \rho_{ge} & \rho_{gg}  
\end{bmatrix} = -\Gamma \begin{bmatrix}
    \rho_{ee} & \frac{1}{2}\rho_{eg} \\
    \frac{1}{2}\rho_{ge} & -\sum_q \widehat{L_q} \rho_{gg} \widehat{L_q}^\dag
\end{bmatrix},
\end{equation}
where we have omitted the primes for simplicity. Now we approximate the equation through adiabatic elimination. Specifically, we eliminate the excited state dynamics assuming $\rho_{ee} = 0$ and solve for $\rho_{eg}$. Then, we rewrite $\dot{\rho}_{gg}$ solely as an equation of $\rho_{gg}$. This provides us with the effective ground state dynamics:

\begin{equation}
\label{equ:gs_dynamics}
\dot{\rho}_{gg} = -i\left[\widehat{H}_{\mathrm{eff}}, \rho_{gg}\right] - \frac{\Gamma_{\mathrm{eff}}}{2} \left(2 \widehat{L_{\mathrm{eff},q}} \rho \widehat{L_{\mathrm{eff},q}}^\dag - \left\{\widehat{\Lambda}, \rho\right\}\right),
\end{equation}

\noindent where we have defined $\widehat{H_{\mathrm{eff}}} = \delta_{\mathrm{eff}}\widehat{\Lambda}$, with $\widehat{G}^- \widehat{G}^+ = \widehat{\Lambda}$ and the parameters $\delta_{\mathrm{eff}} = \frac{\delta}{\delta^2 + \Gamma^2/4}$ and $\Gamma_{\mathrm{eff}} = \frac{\Gamma}{\delta^2 + \Gamma^2/4}$ for convenience.  
Moreover, the effective jump operators are given by $\widehat{L_{\mathrm{eff},q}} = \widehat{L_q} \widehat{G}^+$, which signifies first an absorption followed by a spontaneous emission. The set of effective jump operators satisfies the completeness relation:
\begin{equation}
\sum_q \widehat{L_{\mathrm{eff},q}}^\dag \widehat{L_{\mathrm{eff},q}} = \widehat{G}^- \widehat{G}^+ = \widehat{\Lambda}
\end{equation}

\noindent From $\widehat{H}_{\mathrm{eff}}$, we can read off the effective internal potential and transition rates Eq. \ref{equ:eff_internal_potential}
 and Eq. \ref{equ:eff_transition_rate}.
As the ion moves in the harmonic trap, it experiences an AC Stark-shifted potential that depends on its internal state \( |\pm\rangle = |S_{1/2}, \pm 1/2 \rangle \).
The spatial dependence of ion energy is described by the state dependence of the AC stark shift from the polarization gradient and the trapping potential. The ion energy is given by the expression:
\begin{equation}
    U_\pm=\frac{1}{3} s \delta \hbar \left( 1 \mp \sin(2(\phi + k z)) \right) + U_{\text{trap}}
    \label{equ:eff_internal_potential}
\end{equation}
In this expression, \(\delta\) is the laser detuning, $s$ is the saturation parameter, and \(k\) is the wave number. Finally $U_{\text{trap}} = \hbar\omega(\langle n \rangle +1/2)$ is the trap potential.

\begin{figure*}[htbp]
\centering
\includegraphics[width=.9
\textwidth]{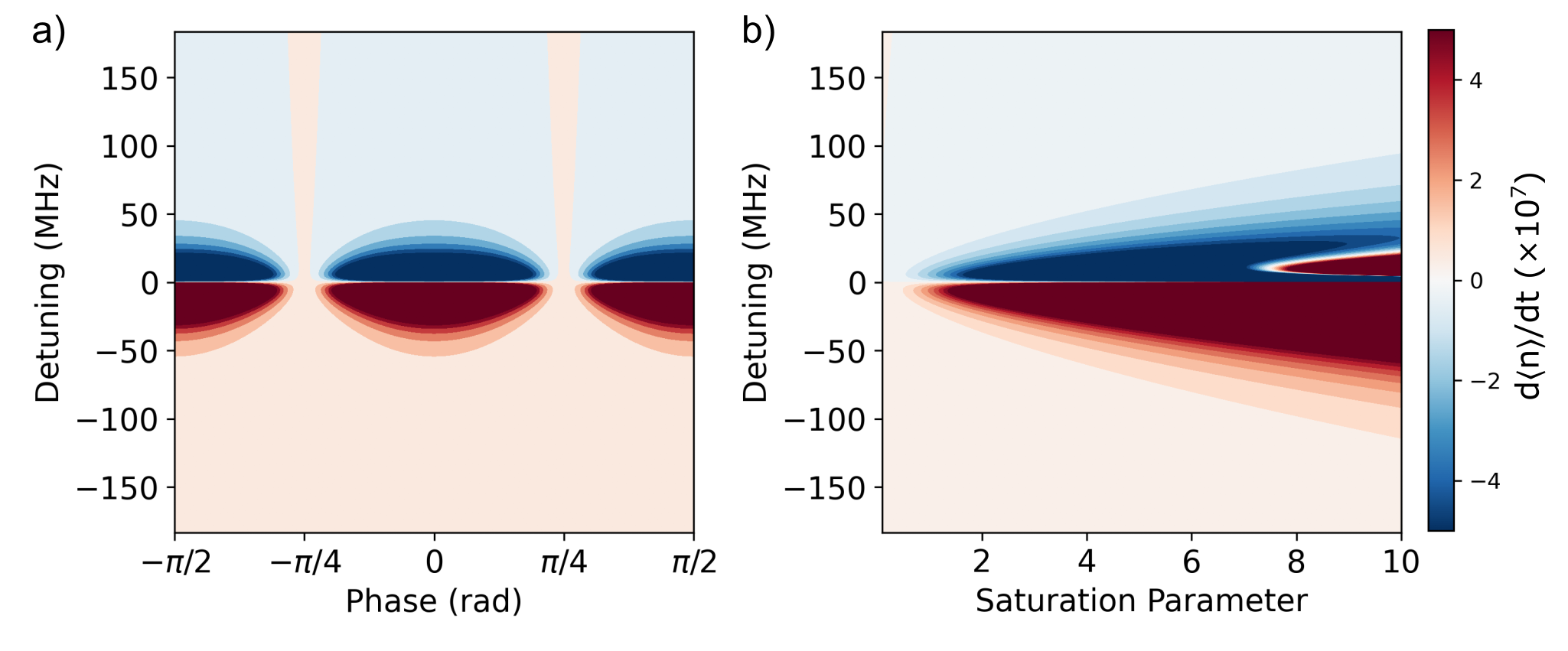}
\caption{Analytical modeling of polarization gradient cooling parameter space. The plots show the frequency/phase (a) and frequency/saturation (b) dependence of the cooling rate determined using the semi-classical approximation for cooling with phase stable polarization gradient cooling. The structure for these plots is much more simplified than the QuTip simulations but does provide a good comparison and verification of the simulated approach.}
\label{fig:analytical_sims}
\end{figure*}

\end{widetext}

\paragraph{Cooling rate and steady state limit}
% \textbf{Summary discussion of the primary heating rates that limit the system relating to photon scattering}

From the second part of Eq.~\ref{equ:effective_gs_dynamics}, we can derive the effective pumping rates between the two ground states. The terms $\Gamma_{\pm\to\mp}$ represent the transition coefficients between these states:

\begin{equation}
    \Gamma_{\pm\to\mp}=
\frac{1}{9} s\Gamma\left( 1 \mp \sin(2(\phi + k z)) \right).
\label{equ:eff_transition_rate}
\end{equation}
With the combination of these two effects, we can approximately compute the cooling rate based on the steady-state value of density matrix $\rho$. We will approximate \( U_\pm \) by expanding it to second order in \( z \). Assuming \( t > \Gamma^{-1} \), the kinetic energy can be described by the equation:
\begin{align}
\dot{E} = \rho_{SS+}\langle\Gamma_{+ \to - }(U_- - U_+)\rangle_+ + \notag \\ \langle\Gamma_{- \to + }(U_+ - U_-)\rangle_- + R.
\end{align}
The steady state solution to the density matrix containing the $S_{1/2}$ ground states is obtained by setting $\dot{\rho}_{gg}$ to $0$ (cf. Eq.~\ref{equ:gs_dynamics}), which gives:

\begin{equation}
\rho_{SS\pm} = \frac{1}{2} \left( 1 \pm \sin\left[ 2(kz + \phi) \right] \right)
\end{equation}
Moreover, \( R \) is the diffusion constant due to the fluctuations of radiation pressure and optical dipole forces arising from the laser-ion interaction~\cite{gordon1980motion}. Similar to the setup in~\cite{joshi2020polarization}, we take $R=H_{\text sc}\hbar\omega$ where $H_{\mathrm{\text sc}}=H_{\mathrm{carr}}+H_{\mathrm{\text sb}}$, the sum of the heating rates for the carrier transition and sideband transitions, which are given by the following equations:

\begin{align}
H_{\text{carr}} =& \frac{\alpha}{3} \eta^2 \Gamma s (1 - \sin^2(2\phi)),\\
H_{\text{sb}} =& \frac{1}{3} \eta^2 \Gamma s (1 + \sin^2(2\phi)),
\end{align}
where \(\eta = \sqrt{\frac{\hbar k^2}{4m\omega}}\) is the Lamb-Dicke parameter, and \(\alpha = \frac{1}{3}\) results from the spatially isotropic spontaneous emission on the \(S_{1/2} \leftrightarrow P_{1/2}\) transition that we are considering. In Ref.~\cite{joshi2020polarization} a dimensionless parameter \(\xi = \frac{\delta s_0}{3\omega_z(1+4\delta^2/\Gamma^2)} \) is defined using the detuning (\(\delta\)), natural linewidth (\(\Gamma\)), axial mode frequency (\(\omega_z\)), and resonant saturation parameter (\(s_0\)) that characterizes the cooling dynamics. This provides a more direct way to parameterize this problem. From here, we can work out the cooling rate $W [\hbar\omega/s]$ and limiting energy, defined by the equation
$\dot{E}=-W(E-E_0):$
\begin{equation}
\begin{aligned}
W(\phi) &= \frac{16}{9} \eta^2 \Gamma s \xi \cos^2 (2\phi), \\
H(\phi) &= \frac{2}{9} \eta^2 \Gamma s \left( 8 \xi^2 \cos^4 (2\phi) + 2 + \sin^2 (2\phi) \right).
\end{aligned}
\end{equation}

\noindent Plots illustrating the cooling rates over the same parameter range as Fig.~S\ref{fig:model_overview} for this equation are plotted in Fig.~S\ref{fig:analytical_sims} as a comparison.

\bibliographystyle{apsrev4-2}
\bibliography{supplemental}% Produces the bibliography via BibTeX.